\documentclass[twocolumn, trackchanges]{aastex7}
\usepackage{graphicx}

\begin{document}

\title{Tracing Fe K X-ray reverberation lag in the energy-resolved spectra of Narrow-line Seyfert 1 galaxy Ton\,S180}

\author[]{Dilip Kumar Roy}
\affiliation{Department of Physics, Rabindranath Tagore University, Hojai, 782435, Assam, India}
\email[]{}  

\author[]{Samuzal Barua}
\altaffiliation{}
\affiliation{Shanghai Astronomical Observatory, Chinese Academy of Sciences, 80 Nandan Road, Shanghai 200030, China}
\email[]{samuzal.barua@gmail.com}
\correspondingauthor{Samuzal Barua}

\author[]{Ranjeev Misra}
\affiliation{Inter-University Centre for Astronomy and Astrophysics (IUCAA), PB No. 4, Ganeshkhind, Pune, 411007, India}
\email[]{}

\author[]{Rathin Sarma}
\affiliation{Department of Physics, Rabindranath Tagore University, Hojai, 782435, Assam, India}
\email[]{}

\author[]{V. Jithesh}
\affiliation{Department of Physics and Electronics, Christ University, Hosur Main Road, Bengaluru 560029, India}
\email[]{}

\begin{abstract}

We report the Fe K relativistic reverberation feature for the first time in the Narrow-line Seyfert\,1 galaxy Ton\,S180. Using a long observation from {\it XMM-Newton} we find that the Fe K emission lag peaks at $117\pm49$ s in the lag energy spectrum computed for frequencies $(0.3-8.5) \times 10^{-4}$ Hz. The lag amplitude drops to $22.85\pm14.20$ s as the frequency increases to $(8.5-30) \times 10^{-4}$ Hz. The time-averaged spectrum of the source shows a relatively narrow Fe K line at $\sim6.4$ keV, resulting in black hole spin to be low ($a=\rm 0.43_{-0.14}^{+0.10}$) found from the reflection modelling. We perform general relativistic transfer function modelling of the lag energy spectra individually. 
This provides an independent timing-based measure of the spin at $a=0.30_{-0.17}^{+0.34}$, and black hole mass $M_{\rm BH} = 0.29_{-0.16}^{+0.01}\times10^8M_{\odot}$, comparable to the previous measurement, and height of the corona $h = 2.59_{-0.33}^{+5.17}r_{\rm g}$. Further, we observe that the Fe K lag and the black hole mass fit well in the linear lag-mass relation shown by other Seyfert 1 galaxies.

\end{abstract}



\section{Introduction}

Accretion of matter onto the supermassive black hole plays a key role in the formation and evolution of galaxies. It is known that most X-ray emissions emerge close to the black hole, where the accretion flow becomes progressively active, and major physical phenomena occur. This central region is generally too small to be spatially resolved by the current detectors, except in a few unusual cases.
X-ray reverberation approach therefore emerged as a powerful technique that overcomes this limit by using echoes of light, providing insight into the structure and kinematics of matter flowing onto the black hole as well as the X-ray emitting corona \citep[see][for a review]{Uttley2014, Cackett2021}.

\begin{figure*}
\includegraphics[width=\textwidth]{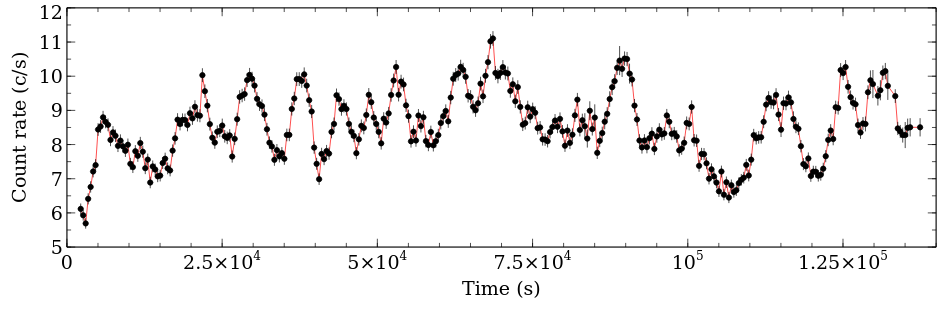}
\vspace{-0.7cm}
\caption{{\it XMM-Newton} lightcurve of Ton\,S180 with a 400 s bin size. The lightcurve is extracted from 0.3 to 10 keV.}
\label{lcurve_400s}
\end{figure*}

The interplay of the accretion disc and the corona is known to lead to X-ray emission in AGN. In the standard paradigm, the hard X-ray emission in AGN is a result of Compton up-scattering of the soft photons from the geometrically thin and optically thick accretion disc \citep{Shakura1973} by the hot electrons in the corona \citep{Haardt1991}. These X-rays are generally referred to as primary continuum described by the powerlaw, a fraction of which focuses on the accretion disc and is reprocessed, producing the reflection spectrum. The pronounced reflection features often appearing in the X-ray spectrum are the fluorescent Fe K emission line at $\sim6.4$ keV and the Compton-hump above $\sim10$ keV \citep{George1991}. The Fe K line profile becomes relativistically broadened due to the combined effect of the gravitational redshift and the Doppler shift \citep{Reynolds2003}, providing information on the black hole spin, disc inclination and the geometry of the inner accretion disc. While other form of the reflection feature is seen in the soft X-ray excess \citep{Crummy2006}, it's origin remains a mystery which in some AGNs has been found to originate from a thermal Comptonization component \citep{Lohfink2013}.

X-ray observations enable us to move beyond spectral modelling and measure the reverberation time delay between the variations of two emission bands. The reverberation time delay is generally attributed to the light travel time between the directly observed continuum from the corona (powerlaw-dominated X-ray band) and the reflected X-rays from the accretion disc \citep{Uttley2014}.

Relativistic reverberation feature was detected for the first time in the Narrow-line Seyfert galaxy 1H0707-495 in which soft X-ray in the 0.3--1 keV band was found to lag the 1--4 keV hard continuum by $\sim30$ s at high frequencies \citep{2009Natur.459..540F}. Follow-up observations have enabled the detection of soft lag in several other Seyfert galaxies and X-ray binaries (XRBs) \citep[e.g.][]{DeMarco2011, Zoghbi2011REJ1034+396, Tripathi2011, Cackett2013, DeMarco2013, Kara2013IRAS13224-3809, Alston2020, Kara2019, Wang2022}. The reported soft lags at high frequencies refer to the light travel time between the soft X-ray excess dominated by relativistic reflection and the powerlaw continuum of the corona. Sources also show hard lags in which the powerlaw-dominated hard X-ray band lags the soft band at low frequencies (a.k.a low-frequency hard lag), which has been observed in  Galactic black holes and AGNs \citep[e.g.][]{Miyamoto1989Natur, Nowak1999a, McHardy2007, Ar2006, ArMcHardy2008}. The possible origin of the hard lag has been interpreted as a consequence of fluctuation rate variation of the accretion flow that propagates over the disc \citep{Ar2006}, also refereed to as propagation fluctuation delay. However, the interpretation of the low-frequency lag scenario remains incomplete.

An alternative model was suggested by \citet{Miller2010a} to account for the origin of low-frequency hard lag, according to which it is the result of reflection from the far side of the disc, typically from a few $\sim 100 R_{\rm g}$ to $\sim 1000 R_{\rm g}$. Timing analysis was carried out by the authors, reporting the hard lag in the Narrow-line Seyfert 1 galaxy in NGC 4051. The same model was suggested to explain the high-frequency lag measured in 1H0707-495 \citep{Miller2010b}, which, contrasts, however, with the soft-lag interpretation of \citet{2009Natur.459..540F}. The follow-up study of \citet{Zoghbi20111H0707} on the same source 1H0707-495, further argued that the inner disc reflector is responsible for reverberation origin in the soft X-ray band, as observed in \citet{2009Natur.459..540F}.

It has also been revealed that the high-frequency lag is not the only reverberation origin of the soft X-ray band; a clear delayed response in the Fe K emission band appears mostly in supermassive black hole candidates, typically the Seyfert 1 galaxies. This has been interpreted to be a consequence of reverberation from within a few gravitational radii of the black hole. The first Fe K reverberation feature was detected in the bright Narrow-line Seyfert 1 galaxy NGC\,4151 by \citet{Zoghbi2012NGC4151}, suggesting the delayed response of the red wing of the broad Fe K profile with respect to the continuum variation. Successive campaigns have explored the underlying feature in several sources \citep[e.g.][]{Zoghbi2013FeKlag, Kara2013IRAS13224-3809, Marinucci2014, Alston2015, Kara2016FeKlag}. 

\begin{figure}
\centering
\vspace{0.35cm}
\includegraphics[width=0.85\columnwidth]{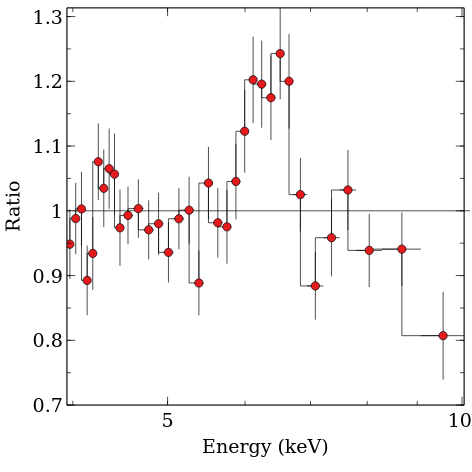}
\caption{Resolving the Fe K emission line profile, produced from the ratio of the X-ray spectrum to a simple powerlaw model. The data is binned to show only a few points for visual purpose.}
\label{ratio_data_pl}
\end{figure}

The Narrow-line Seyfert\,1 galaxy Ton\,S180 is among the objects that shows the Fe K emission feature, making it a plausible candidate for exploring the reverberation lag in the corresponding energy band. A broadband spectral analysis was conducted by \citet{Walton2013TonS180} using a long (100-120 ks) {\it Suzaku} observation taken in 2006. The authors successfully reproduced the X-ray spectrum by modeling with relativistic reflection. Their analysis suggested a rapidly rotating black hole in {Ton\,S180}, with spin constrained at $0.91_{-0.09}^{+0.02}$ -- consistent with the scenario in which a black hole grows through prolonged and ordered accretion. The source was found to have little to no intrinsic absorption to complicate the analysis. Reflection spectroscopy was further undertaken by \citet{Parker2018} using the 2015 {\it XMM-Newton} observation. The powerlaw residuals to the spectrum yield a clear Fe K line profile, which appeared to be strong but relatively narrow. The X-ray spectrum of the source cannot be reproduced simply by relativistic reflection -- leaving a smoother soft excess. The spectral fit favoured a low spin value of the black hole, which was constrained to be $a <0.7$. Lastly, \citet{Matzeu2020} broadly explored the reflection dependency in Ton\,S180 using 2000-2016 {\it XMM-Newton} and {\it NuSTAR} observations, suggesting that the broadband X-ray spectrum can be described by the relativistic reflection combined with (a) disc thermal emission, (b) seed photon comptonisation in an optically thick corona (i.e., warm corona; $kT_{\rm e} \sim 0.3$ keV), and (c) Comptonisation in the hot ($kT_{\rm e} \ge  100$ keV) and optically thin corona. Analysis of the spectra with different models provided a black hole spin from low ($<0.34$) to high value ($>0.98$).


Timing analysis is yet another powerful tool that allows for a model-independent approach to examine the peak Fe K line emission feature by measuring its time delay. 
In fact, the Fe K emission band is the clearer part of the spectrum, allowing the search for a reverberation signature. In this work, we search for the Fe K reverberation in Ton\,S180 by computing Fourier lag spectra from one of the long {\it XMM-Newton} observations. We extend the work to reverberation modelling to constrain the key physical parameters, spin and mass of the black hole, disc inclination as well, and the height of the X-ray corona.

The paper is structured as follows: Section 2 presents the observations and data analysis. The timing analysis is described in Section 3. In Section 4 we report on the modelling of the time-averaged spectrum and lag spectra. The results are discussed and interpreted in Section 5.

\begin{figure*}
\centering
\includegraphics[width=0.8\columnwidth]{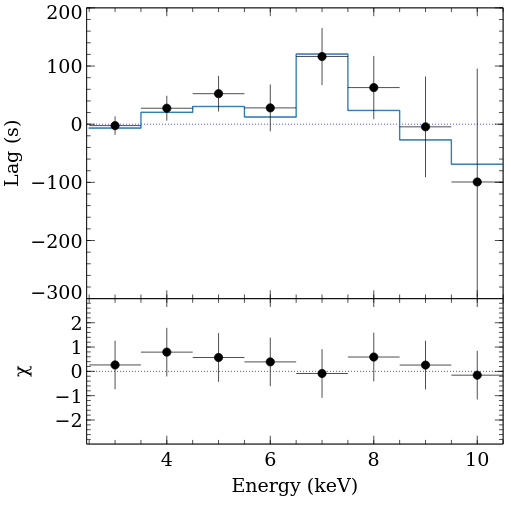}
\hspace{1cm}
\includegraphics[width=0.8\columnwidth]{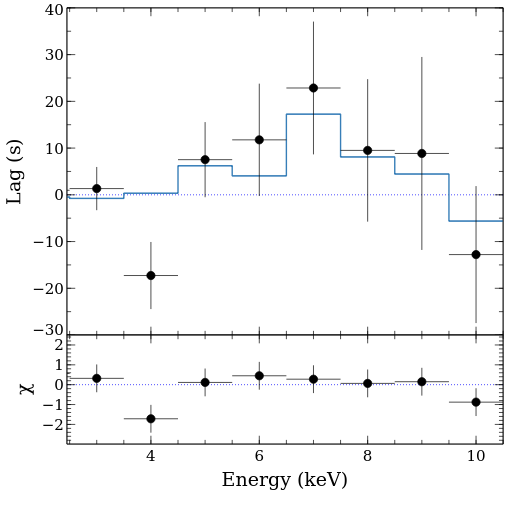}
\caption{Lags as a function of energy for Ton\,S180 computed for frequencies $(0.3-8.5) \times 10^{-4}$ Hz (left panel) and $(8.5-30) \times 10^{-4}$ Hz (right panel). The lag spectra show a clear Fe K emission feature with peak lags of $117\pm49$ s at frequencies $(0.3-8.5) \times 10^{-4}$ Hz and $22.85\pm14.20$ s at frequencies $(8.5-30) \times 10^{-4}$ Hz. The lags are fitted using the general relativistic transfer function model {\tt KYNREVERB}.   
Model lags are represented by the blue steps, whereas the black data points are the computed Fourier lags. Zero lag as a function of energy is shown as the blue dotted line.}
\label{model_lagenergy}
\end{figure*}

\section{Observation and data reduction} \label{section2}

\subsection{\it XMM-Newton}

{\it XMM-Newton} conducted the longest observation of Ton\,S 180 in 2015 July (Obs. ID 0764170101), obtained for a total duration of $\sim141$ ks. 
The lightcurve from this observation is shown in Figure~\ref{lcurve_400s}, which shows significant X-ray variability with a high count rate; both are particularly useful in the measurement of reverberation lag.
We reduced the {\em XMM-Newton} data with the Science Analysis System (sas v.20.0.0). We used the sas task {\sc evselect} to create the event file lists for the EPIC-pn detector. The events were cleaned for high background flaring by applying the filtering condition (PATTERN $<= 4$)\&\&(FLAG == 0). We created a Good Time Interval (GTI) to exclude the background flare above an optimal value. The cleaned events were then used to extract the source and background regions. We selected a circular extraction region of 35\arcsec for the source, and a larger background region of 50\arcsec was selected from the source free space. In the following step, we extracted source and background lightcurves using these region files. Finally, background-subtracted lightcurves were produced using the sas task {\sc epiclccorr}.

\subsection{\it NuSTAR}

{\it NuSTAR} observed Ton\,S180 for an exposure of $\sim127$ ks exposure (Obs. ID 60201057002) in 2016.
We reduced the data using {\it NuSTAR} data analysis software {\sc nustardas}. The data were  screened and processed using the stranded {\sc nupipeline} script, which provides Level-1 events. The events were cleaned and calibrated using the most recent version of the calibration database ({\sc caldb}; v.20250415). We extracted source counts from a circular region of 50\arcsec in radius, while a larger background region of 90\arcsec was used for both the {\it NuSTAR} detectors FPMA and FPMB. The source and background spectra were then produced with the {\sc nuproducts} task. Finally, we grouped the spectra using the {\tt grppha} tool so that each spectral bin contains 50 counts.

\section{Time lag analysis for {Ton\,S180}}

We calculated the lag spectra using the direct Fourier method outlined in \citet{Nowak1999a} and further demonstrated in \citet{Uttley2014}. We produced 8 different lightcurves of finer energy bins ([2--3, 3--4, 4--5, 5--6, 6--7, 7--8, 8--9, 9--10] keV) and took their Fourier transformation. The lightcurves were extracted for a bin size of 10 s. We construct the cross-spectrum by multiplying the Fourier transform of the lightcurves by their complex conjugate, which was then averaged over user-defined frequency bins in logarithmic space. The cross-spectrum generally encodes the phase difference between the lightcurves which can eventually be converted into frequency-dependent time lags $\tau (\nu)$ by dividing by $2\pi \nu$, where $\nu$ is the center of the defined frequency bins. Fourier lag analysis was carried out using the publicly available code pyLag\footnote{https://github.com/wilkinsdr/pyLag}.


\subsection{Lag versus energy}

We started by computing the lag energy spectra at different frequency sets. The long observation and small bin size of the lightcurve allow us to calculate lags on a wider frequency range. We used the sign convention, so that positive lags indicate that the band of interest lags the reference band.  The reference band here was taken to be the entire 2--10 keV band, except for the bands of interest. A wider reference band was used to increase the signal-to-noise ratio and ensure reduction of the correlated noise between the narrow energy bands.


Computing lags versus energy, we observe that the lag spectra trace the peak lag in the Fe K emission band, 6--7 keV,  while the other bands appear to lead relative to this band.
The resulting spectra are shown in Figure\,\ref{model_lagenergy}. There appears to be a sharp drop of the lag amplitudes beyond the 7 keV band, which is typically dominated by the powerlaw emission (directly observed continuum from the corona), while below 4 keV bands are roughly consistent with zero lag. The entire lag spectrum shifts downward with decreasing lags in each energy band as the frequencies increased from $(0.3-8.5) \times 10^{-4}$ Hz (left panel) to $(8.5-30) \times 10^{-4}$ Hz (right panel), retaining the observed spectrum shape where the Fe K peak lag remains. Frequencies of this order are of particular importance, at which the observed lag spectra replicate the shape of a reflection spectrum, with the Fe K emission feature at $\sim6.4$ keV. Interestingly, the shape more mimics the Fe K line profile produced in Figure\,\ref{ratio_data_pl}. 
There appears a dip at 4 keV in the lag energy spectra computed for higher frequencies (Figure\,\ref{model_lagenergy}, right), indicating the earliest response. This feature is more common in the lag energy spectra as a result of relativistic reverberation around the black hole \citep[see also][]{KaraArk564Mrk335, Wilkins2017}. 
We have checked the log-linear trend of the observed lags by using the log-linear model $y=a+b\rm log(x)$, which provides a poor fit with a $\chi^2$ value of 14 for 8 degrees of freedom for the two frequency ranges. This indicates that we can rule out the null hypothesis of a simple log-linear fit and implies that the energy dependent lag traces the Fe K emission feature.

\subsection{Lag versus frequency}

\begin{figure}
\centering
\includegraphics[width=0.75\columnwidth]{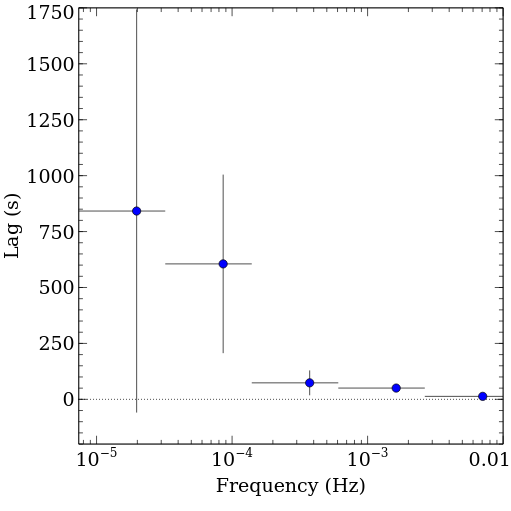}
\caption{Lag frequency spectrum of Ton\,S180 computed between the power law-dominated 2--4 keV band and reflection-dominated 4--7 keV band. Positive lags indicate hard lags (4--7 keV hard band lags the 2--4 keV soft band). 
The lag spectra show a decreasing trend with increasing frequencies, where Fe K emission lag is clear.}
\label{lagfreq_2-3vs5-7keV}
\end{figure}

Frequency-dependent lags were computed between the two X-ray energy bands: one is the reflection-dominated hard band consisting of Fe K emission peak (taken as the band of interest), and the other is the powerlaw-dominated soft band (taken as the reference band). Again, we used the same sign convention, in which hard lags are indicated by positive values. The resulting lag spectrum is shown Figure\,\ref{lagfreq_2-3vs5-7keV}. The lag spectrum shows the evolution of lags as a function of the temporal frequency produced between the reflection-dominated 4--7 keV band and the powerlaw-dominated 2--4 keV band. The observed lags follow a trend with the lag decreasing with increasing frequency similar to other Narrow-line Seyfert galaxies (see e.g. MCG–5-23-16 and NGC\,7314 \citep{Zoghbi2013FeKlag}).
In the lag spectrum, we see that the Fe K band lag is clear up to a cutoff frequency, above which the lags are observed to decrease promptly, mostly consistent with zero, as the lags are dominated by Poisson noise. 

A point to be noted is that since only two narrow energy bands are used in the  lag frequency spectrum, the Poisson noise does not allow us to see the lags at further higher frequencies. The energy-dependent lags are, therefore, more reliable for clearly tracing the peak lag in the Fe K band, where the whole energy range (2--10 keV) was taken as a reference.

\begin{figure}
\centering
\includegraphics[width=0.75\columnwidth]{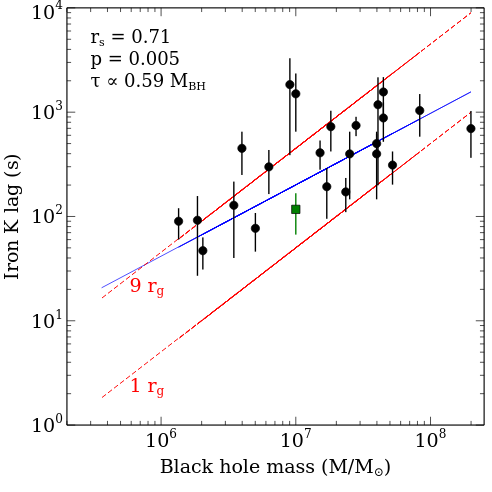}
\caption{Fe K lag amplitudes versus black hole mass. The plot is reproduced from the published results (shown in black) reported in Table\,2 of \citep{Kara2016}, in which our lag measurement in Ton\,S180 (shown in green) is overplotted. The blue line indicates a linear model used to fit the data, providing a scaling relation of the lag with the black hole mass as $\tau \propto 0.59~M_{\rm BH}$. The red diagonal lines indicate lag at $1r_g$ and $9r_g$. The Spearman correlation test provides a Spearman rank correlation coefficient of 0.71 at a probability of 0.005.}
\label{lag_bh_mass}
\end{figure}

\subsection{Scaling Fe K lag with the black hole mass}

The amplitudes of Fe K lags for supermassive black holes have been observed to follow a linear dependence on the black hole mass \citep[see][]{Uttley2014, Kara2016FeKlag}. Fe K emission lag appears to be larger than the soft excess lags \citep{Uttley2014}. Here, we reproduce the plot for Fe K lag versus black hole mass from \citet{Kara2016FeKlag} (obtained from the published results) and include the results for Ton\,S180 obtained in this work. The resulting plot is shown in Figure\,\ref{lag_bh_mass}. We used a linear model to scale the lags as a function of black hole mass, which shows a linear mass dependence of the lags as $\tau \propto 0.59~M_{\rm BH}$. Furthermore, we estimated the Spearman rank correlation coefficient, $r_s = 0.71$ with a probability p = 0.005. These estimates are consistent with the results obtained by \citet{Kara2016FeKlag} who derived the lag-mass relation for the same sample of type 1 Seyferts. In the plot, the red diagonal lines indicate the lags at $1r_g$ and $9r_g$ for different Seyfert\,1 galaxies. When the Fe K lag amplitude measured for Ton\,S180 is included, it lies within this interval, following the linear trend, further confirming the Fe K reverberation for a black hole mass of $2\times10^{7}~M_{\odot}$ (with a factor of 2 uncertainty) \citep{Turner2002}.


 \begin{figure}
\vspace{-2cm}
\hspace{-1.0cm}
\includegraphics[width=1.05\columnwidth, angle=-90]{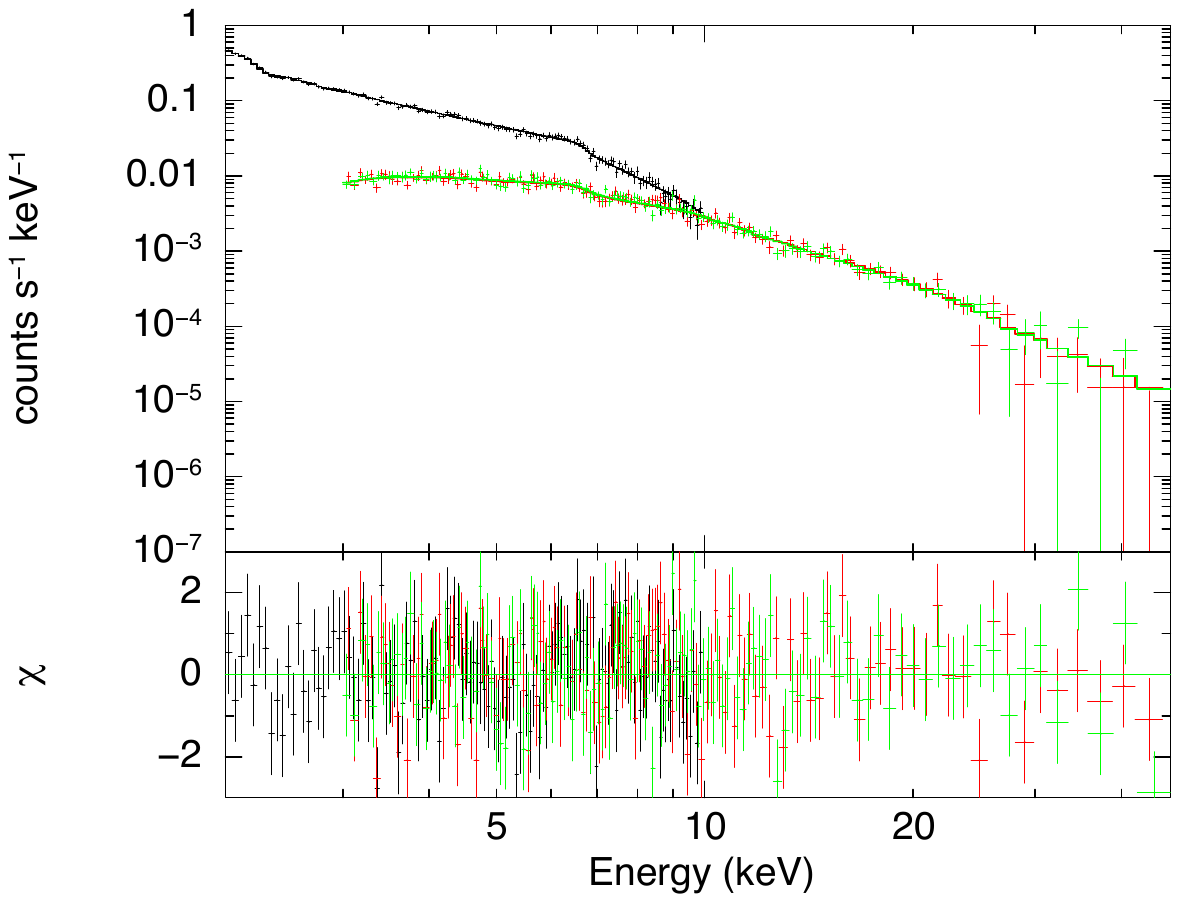}
\vspace{-0.7cm}
\caption{Time-averaged spectra of Ton\,S180 produced from {\it XMM-Newton} and {\it NuSTAR} observations. The spectra are fitted simultaneously using the relativistic reflection model {\tt relxilllp} (see text).}
\label{2-10keV_spectra}
\end{figure}

\section{Modelling}

\subsection{Time-averaged spectrum}

We performed a spectral analysis to account for the reflection dependency of the source. Previous analysis showed that the X-ray spectrum contains a relatively narrow Fe K line \citep{Parker2018, Matzeu2020} as shown in Figure~\ref{ratio_data_pl}, reproduced here from the ratio of the spectrum to a simple powerlaw model. We carried out iron-line spectroscopy to constrain some key physical parameters, with particular focus on the black hole spin.

We therefore fitted the time-averaged spectrum in {\sc xspec} using the {\tt relxilllp} component of the relativistic reflection model {\tt relxill} \citep{Garcia2014, Dauser2014}, which convolves the X-ray spectrum being due to reflection from the disc as a result of irradiation from the X-ray source, described by {\tt xillver} \citep{Garcia2013}, with the relativistic ray-tracing code {\tt relline} \citep{Dauser2013} which accounts for the gravitational redshifts and Doppler shifts that cause relativistic broadening. We used the {\tt Tbabs} model \citep{Wilms2000} to take into account the galactic absorption. The fit was performed over a broadband spectral coverage inclusive of {\it XMM-Newton} observation simultaneously with that from {\it NuSTAR}. Combining the {\it NuSTAR} observation with {\it XMM-Newton} the spectral resolution can be maximized together with the photon count across 3-50 keV. We used a cross-calibration constant between the two instruments, fixing at 1 for {\it XMM-Newton} while left free for {\it NuSTAR}. The soft X-rays in Ton\,S180 are  contributed by multiple components, including thermal emission from the accretion disc, Comptonized emission from the warm and optically thick corona. While describing the broadband spectrum by the reflection model, there occurs a mismatch between the fits of the thermal emission and reflection, as both are responsible for different physical processes, causing an inappropriate measurement of the black hole spin \citep{Parker2018, Matzeu2020}. Therefore, to place a cleanest measurement on the reflection spectrum and hence the spin, we fitted the spectrum over the 2--50 keV band \citep[see also][for example]{Zoghbi2012NGC4151, Wilkins2021Natur}. 

The relativistic reflection model {\tt relxilllp} provided a good fit to the broadband spectrum, providing a reduced $\chi^2/d.o.f.=366/380$. Our simultaneous fit to the {\it XMM-Newton} and {\it NuSTAR} spectra provided constraints on the key physical parameters, the photon index $\Gamma$, iron abundance $A_{\rm fe}$, coronal height $h$, and the ionization parameter log$\xi$. We obtained the black hole spin at a low value $a=0.43_{-0.14}^{+0.10}$ at an inclination of $42.72_{-1.42}^{+1.51}$ degrees. The reflection fraction has been found to be $R=1.54_{-0.36}^{+0.27}$, which is moderately high, implying a strong reflection dependency of the source. During the fit, we fixed the inner radius at $1r_{\rm g}$, the outer radius at $400r_{\rm g}$, and the emissivity indices at the classical value of $q=3$ \citep[see][]{Matzeu2020}. The best-fit parameters are shown in Table~\ref{spectral_results}. We show the modelled spectra from {\it XMM-Newton} and {\it NuSTAR} in Figure~\ref{2-10keV_spectra}.

\begin{figure}
\centering
\includegraphics[width=1.0\columnwidth]{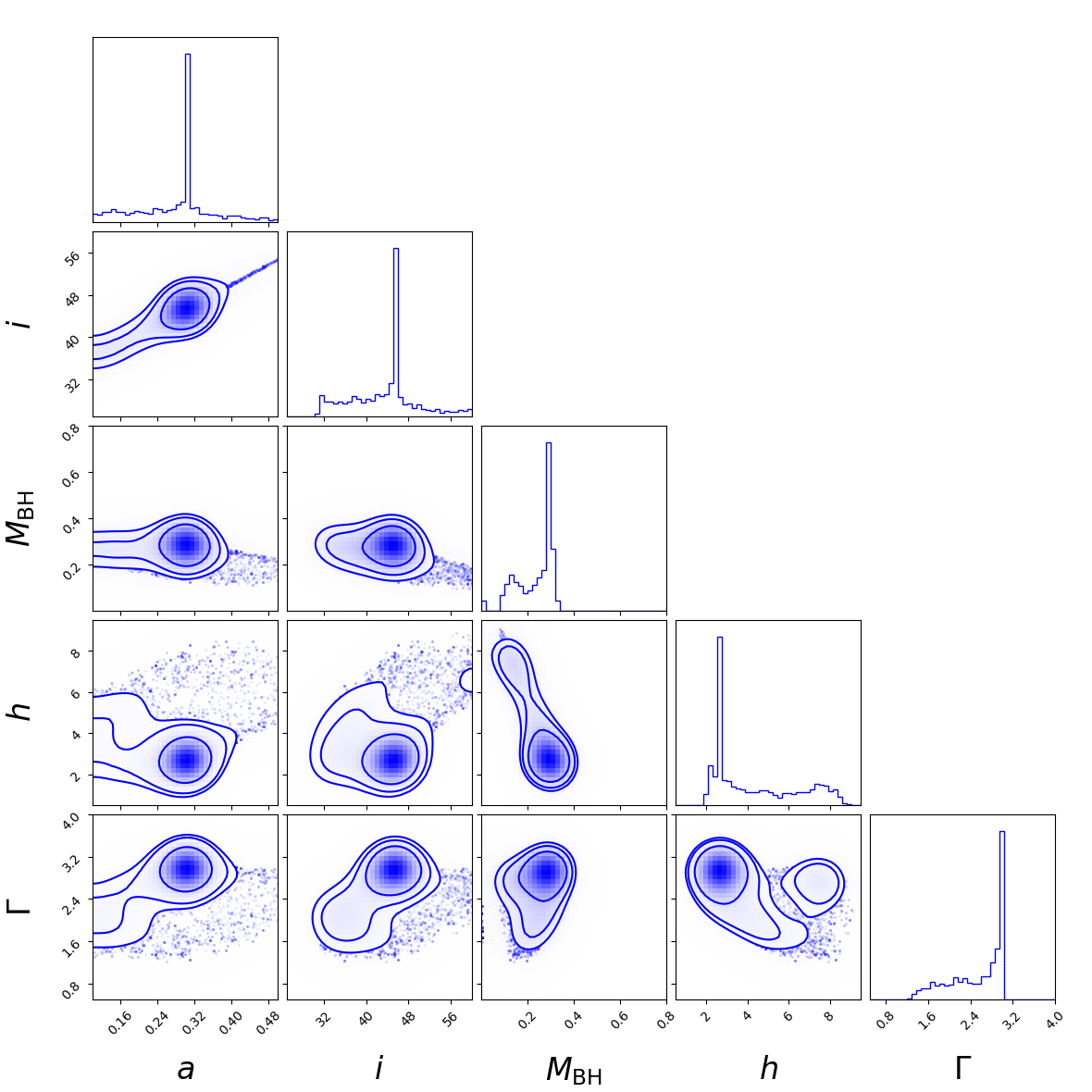}
\caption{Corner plots of the MCMC parameters obtained from the relativistic transfer function modeling. The one-dimensional histograms represent the posterior probability distribution, which are normalized to have total area of one. Shown are the black hole spin $a$, inclination $i ~(\theta^\circ$), black hole mass $M_{\rm BH}~(\times10^8M_\odot$), coronal height $h~(r_{\rm g})$ and photon index $\Gamma$.}
\label{cornerplot}
\end{figure}

\begin{table*}
	\centering
  \caption{Parameters obtained from the fits of {\it XMM-Newton} and {\it NuSTAR} spectra. All parameters are tied betwen the two detections. During analysis the multiplying constant is fixed at 1 for {\it XMM-Newton} and left free for {\it NuSTAR}. Galactic absorption is fixed at $N_H=1.3\times10^{22}~\rm cm^{-2}$. The 2--10 keV luminosity is obtained to be $3.16\times10^{43}~\rm erg~s^{-1}$. }
	\label{spectral_results}
	\setlength{\tabcolsep}{3.45pt}
	\begin{tabular}{lcccccccccc} 
		\hline
		  $N_{\rm H}$ & $\Gamma$  & $a$ & $i$ & log$\xi$ & $A_{fe}$ & $ R$ & $h$ & $\chi^2/d.o.f$\\

		         &          &  (${GM}/c^2$)   & ($\rm degree$) & log(erg cm $\rm s^{-1}$) & Fe/solar &    & $r_{\rm g}$ & \\
		
		\hline
		
		   $1.3\times10^{22}~\rm cm^{-2} (f)$ &  $2.32_{-0.02}^{+0.02}$ & $0.43_{-0.14}^{+0.10}$ & $42.72_{-1.42}^{+1.51}$ & $2.65_{-0.28}^{+0.10}$ & $\rm 1.29_{-0.26}^{+0.38}$ & $1.54_{-0.36}^{+0.27}$ & $ 2.57_{-0.31}^{+0.44}$ & 366/380 \\\\
			
		\hline
	\end{tabular}
\end{table*}

\subsection{lag-energy spectrum}

The intrinsic powerlaw continuum arising from the corona and the reflection-dominated emission from the accretion disc are generally expected to be observed with a certain time delay, as seen in the lag frequency or lag energy spectra. However, in reality, the flux of both emissions contributes to each band, diluting the absolute value of the measured lag, known as the lag dilution effect \citep[see also][]{Uttley2014}. It is difficult to separate out the components of the direct continuum and the reflected emission because of the lack of high-cadence observations with the current detectors. This is generally the case for optical reverberation mapping, where lightcurves from low and high-cadence observations can be used to separate out lags due to two physical processes. 
Even in Fourier space, X-ray/UV/Optical reverberation mapping has recently used reverberation modeling to decouple the reprocessing lag and the BLR continuum lag \citep[e.g.][]{Cackett2022}.

X-ray reverberation modeling is yet another powerful approach that has been performed in Fourier space. We used the relativistic transfer function model {\tt kynreverb}\footnote{https://projects.asu.cas.cz/stronggravity/kynreverb} \citep{Caballero-Garc2018} to the observed lag-energy spectra, in which Fe K lag peaks (see Figure\,\ref{model_lagenergy}). The model relies on the full treatment of the general relativistic effect around an accreting black hole, allowing one to measure the path length difference between the disc and the corona to the rest frame observer. Disc reflections from each radius of the ionised disc are estimated from the reflection table {\tt xillver} \citep{Garcia2013, Dauser2014} or {\tt reflionx} \citep{Ross2005}. We fitted the two lag energy spectra simultaneously in {\sc xspec}. In the fit, a few parameters were fixed to the values obtained from our fits of the time-averaged spectrum, inncluding the iron abundance ($A_{\rm fe}$) and 2--10 keV luminosity at $3.16\times10^{43}~\rm erg~s^{-1}$. 
Additionally, we fixed the inner radius ($R_{\rm in}$) at $1r_g$, and the outer radius ($R_{\rm out}$) at $400r_g$ similar to the time-averaged spectrum (see Section 4.1). The photon index ($\Gamma$), black hole spin ($a$), coronal height ($h$), mass of the black hole ($M_{\rm BH}$), disc inclination ($i$) are left free to vary independently, which have been constrained at $\Gamma >1.88$, $a =0.30_{-0.17}^{+0.34}\,{GM}/c^2$, $h = 2.59_{-0.33}^{+5.17}r_{\rm g}$, $M_{\rm BH} = 0.29_{-0.16}^{+0.01}\times10^8M_{\odot}$ and $i = 45.40_{-7.82}^{+16.81\,\circ}$. The fit provided the electron density of the disc at $n_{\rm d} = 16.04_{-0.52}^{+1.20}\times10^{15}~\rm cm^{-3}$, consistent with the previous measurements $15.6~\rm cm^{-3}$ \citep{Jiang2019} and $<16~\rm cm^{-3}$\citep{Matzeu2020}. The parameters provided by the lag spectra are found to be within the uncertainties of those obtained from the flux spectrum, in particular, both fits provided a relatively lower value of the black hole spin. The timing-based parameters are listed in Table\,\ref{response_model_lagenergy}. The best-fit modelling provided a reduced $\chi^2/d.o.f. =16/12$. We present the best-fit modelled spectra in Figure\,\ref{model_lagenergy}.

The errors on the parameters obtained from the time-averaged spectrum and the lag spectra were estimated at 68.27\% confidence by performing a Markov Chain Monte Carlo (MCMC) analysis. We used the Jermy Sandars {\sc xspec\_emcee}\footnote{https://github.com/jeremysanders/xspec\_emcee} code in {\tt xspec}, which is a pure Python implementation of the Goodman \& Weare's Affine Invariant MCMC Ensemble sampler. For analysis, we used 50 walkers with 10,000 iterations and burned the first 1000 from the MCMC chain. We produce the corners of the MCMC parameters obtained from the transfer function modeling and plot their posterior distributions in Figure\,\ref{cornerplot}. 

\begin{table}
	\centering
	\caption{Best fit parameters from the transfer function modeling. The errors have been estimated at 68.27\% confidence interval. The parameters are photon index $\Gamma$, inclination $i$, coronal height $h$, black hole mass $M_{\rm BH}$,  and spin $a$.}
	\label{response_model_lagenergy}
    \setlength{\tabcolsep}{4pt}
	\begin{tabular}{lccccr} 
		\hline
		$\Gamma$ & $i$ & $h$ & $M_{\rm BH}$ & $a$ \\
                     & ({$\rm degree$})& ($r_g$) & ($10^7M_{\odot}$)& (${GM}/c^2$) &\\
		\hline
		$>1.88$ & $45.40_{-7.82}^{+16.81}$ & $2.59_{-0.33}^{+5.17}$ & $0.29_{-0.16}^{+0.01}$ & $0.30_{-0.17}^{+0.34}$\\
		\hline
	\end{tabular}
\end{table}


\section{Discussion}
We have detected the reverberation of the Fe K emission line for the first time in the Narrow-line Seyfert\,1 galaxy Ton\,S180. 
Generally, the Fe K emission feature appears as a clean part of the X-ray band where absorptions and other effects are less, providing an opportunity to place a robust measurement of the reverberation delay. 
We find that Fe K lag peaks at $\sim117$ s, calculated for the frequency range $(0.3-8.5)\times10^{-4}$ Hz. The lag drops significantly as the frequencies are increased, allowing us to observe the lag amplitude of $\sim22$ s at the highest Fourier frequencies of $(8.5-30)\times10^{-4}$ Hz -- suggesting that the underlying emission arises closer to the black hole.

Fe K emission line in Ton\,S180 has been confirmed in its X-ray spectrum. Using Fe K line spectroscopy, the black hole spin was constrained to $\sim0.98$, obtained by \cite{Walton2013TonS180}. Later observation with {\it XMM-Newton} ($\sim141$ ks duration) showed that the emission feature is relatively narrow, which in turn favours the low spin value ($<0.4$) constrained from the spectral modelling \citep{Parker2018}. The 0.3--10 keV X-ray spectrum cannot be accounted for entirely reflection dominated where the soft X-ray band requires fitting for Comptonisation of the seed photons and the thermal emission. The recent analysis of broadband spectra covering the {\it XMM-Newton} and {\it NuSTAR} bands has shown similar spectral properties, providing a low spin value ($<0.34$) \citep{Matzeu2020}.
However, reverberation modelling using a relativistic transfer function is another effective approach, which provides a timing-based result, allowing for an independent measure on the key physical parameters, spin and mass of the black hole \citep[see e.g.][]{Alston2020}. However, this is generally applicable in sources showing reverberating X-ray features. Our reverberation modelling has put constraints on the black hole spin at $a=0.30_{-0.17}^{+0.34}$ -- consistent with the value from the spectral fit, $a=0.43_{-0.14}^{+0.10}$. We have been able to constrain the black hole mass at $0.29_{-0.16}^{+0.01}\times10^{8}~M_{\odot}$, the uncertainties of which are comparable to the previous measurement by \citet{Turner2002}. Furthermore, it it has been found that the other parameters from the lag spectra are within error bars of those obtained from the spectral fit.

We notice that there appears degeneracy between the black hole mass $M_{\rm BH}$ and the height $h$ of the corona measured from the reverberation modelling for Ton\,S180. 
Inherent degeneracy between the height and black hole mass is not unusual, as it has been seen in previous reverberation modelling results, showing a static picture \citep[e.g.][]{Emmanoulopoulos2011, Chainakun2016, Ingram2019}. However, this could possibly be overcome by performing joint modelling of multiple spectra obtained from numerous observations. Simultaneous modelling of the lag spectra from 16 long {\it XMM-Newton} observations ($\sim130$ ks per orbit) breaks this inherent degeneracy, providing a significant constraint on the black hole mass and coronal height \citep[see][]{Alston2020}. 
%
Spectral fitting results also pointed out that degeneracy also persists between black hole spin $a$ and disc inclination $i$ \citep[see][Figure\,3]{Parker2018}, where above the $3\sigma$ confidence level the upper limit of the spin is insignificant and rules out the maximal rotation of the black hole. The degeneracy appears even at $1\sigma$ confidence. Even using reverberation modelling, we have not been able to break the degeneracy. 
In future multi-epoch observations will likely to help break the degeneracies.


Ton\,S180 has been observed in four different epochs by {\it XMM-Newton} of which only the longer observation with $\sim141$ ks exposure obtained in 2015 allowed us to detect the reverberation feature for the first time in this source. 
In future, a complete picture on the coronal geometry and dynamics of the inner accretion flow can be explored using multi-epoch observations with quite long exposures. Understanding coronal geometry also becomes feasible with the {\it IXPE} \citep{Weisskopf2022}, but it has some limitations, so far observing only a few bright Seyfert galaxies. The upcoming {\it eXTP} \citep{Zhou2025} with four times effective area and better sensitivity than {\it IXPE} and {\it NewAthena} \citep{Cruise2025} are expected to allow robust spectral-timing and polarimetric analysis to provide detailed information on the geometry of the accretion flow and the nature of the corona.

\begin{acknowledgments}
 We thank the anonymous referee for the constructive comments that have improved the manuscript.
This research has made use of data and software provided by the High Energy Astrophysics Science Archive Research Center (HEASARC), which is a service of the Astrophysics Science Division at NASA/GSFC.
DKR acknowledges IUCAA, Pune for the support in this work.
SB acknowledges support from China Postdoctoral Science Foundation General Fund (grant no. 404985), Shanghai Postdoctoral Excellence Program (grant no. 2024686), and National Foreign Expert Project, Ministry of Science and Technology (grant no. 20240238). RS acknowledges the associateship program of IUCAA, Pune. VJ acknowledges the support provided by the Department of Science and Technology (DST) under the ‘Fund for Improvement of S \& T Infrastructure (FIST)’ program (SR/FST/PS-I/2022/208). VJ also thanks the Inter-University Centre for Astronomy and Astrophysics (IUCAA), Pune, India, for the Visiting Associateship.

\end{acknowledgments}

%
\facilities{\it XMM-Newton}

\software{Astropy \citep{2013A&A...558A..33A,2018AJ....156..123A}, numpy, scipy, matplotlib} 





\bibliography{Reference}{}
\bibliographystyle{aasjournalv7}



\end{document}